\documentclass[12pt]{iopart}

\usepackage{cite}
\usepackage{graphicx}
\usepackage{caption}
\usepackage[caption=false]{subfig}
\usepackage{amssymb}
\usepackage{physics}
\usepackage{float}

\usepackage{tikz}

\newcommand{\tikzcircle}[2][red,fill=red]{\tikz[baseline=-0.5ex]\draw[#1,radius=#2] (0,0) circle ;}
\definecolor{green2}{RGB}{0,128,0}

\begin{document}

\title{Practical Limits of Error Correction for Quantum Metrology}

\author{Nathan Shettell$^{1,2}$, William J. Munro$^{3,2}$, Damian Markham$^{1,4}$ and Kae Nemoto$^{2,4}$}

\vspace{7pt}
\address{$^1$ Laboratoire d'Informatique de Paris 6, CNRS, Sorbonne Université, 4 place Jussieu, 75005 Paris, France}
\vspace{4pt}
\address{$^2$ National Institute of Informatics, 2-1-2 Hitotsubashi, Chiyoda-ku, Tokyo 101-8430, Japan}
\vspace{4pt}
\address{$^3$ NTT Basic Research Laboratories \& Research Center for Theoretical Quantum Physics,3-1 Morinosato-Wakamiya, Atsugi, Kanagawa, 243-0198, Japan}
\vspace{4pt}
\address{$^4$ Japanese-French Laboratory for Informatics, CNRS, National Institute for Informatics, 2-1-2 Hitotsubashi, Chiyoda-ku, Tokyo 101-8430, Japan}
\vspace{7pt}
\ead{nathan.shettell@lip6.fr}
\vspace{7pt}
\address{\textbf{Keywords:} Quantum Metrology, Quantum Error Correction, Heisenberg Scaling}
\begin{abstract}
    Noise is the greatest obstacle in quantum metrology that limits it achievable precision and sensitivity. There are many techniques to mitigate the effect of noise, but this can never be done completely. One commonly proposed technique is to repeatedly apply quantum error correction. Unfortunately, the required repetition frequency needed to recover the Heisenberg limit is unachievable with the existing quantum technologies. In this article we explore the discrete application of quantum error correction with current technological limitations in mind. We establish that quantum error correction can be beneficial and highlight the factors which need to be improved so one can reliably reach the Heisenberg limit level precision.
\end{abstract}

\section{Introduction}

Quantum metrology is an auspicious discipline of science where one makes high precision estimates of unknown parameters \cite{giovannetti2004, giovannetti2006, giovannetti2011, toth2014, degen2017}. The field has numerous applications, including spectroscopy \cite{leibfried2004, schmitt2017}, magnetometry \cite{brask2015, taylor2008, wasilewski2010, razzoli2019}, thermometry \cite{neumann2013, correa2015} and gravimetry \cite{qvarfort2018, kritsotakis2018}. It has been known that utilizing quantum resources including entanglement  allows one to achieve a gain in precision when compared to classical strategies \cite{giovannetti2004, caves1981, huelga1997, munro2002, krischek2011, hyllus2012, pezze2014, pezze2018}. In fact quantum resources allows one to reach a level of precision not possible in the classical situation \cite{giovannetti2006, bollinger1996, hyllus2012, pezze2014}. There is an ultimate precision limit enabled by quantum mechanics called the Heisenberg limit \cite{yurke1986, holland1993, giovannetti2006} which can only be achieved under idealized conditions. However, in realistic experiments, noise and experimental imperfections limit ones achievable precision and sensitivity \cite{escher2011, demkowicz2012, kolodynski2013, chaves2013, haase2016}. In fact such noise can, in principle, limit ones sensitivity to that achievable by classical approaches \cite{chin2012}.

Quantum error correction \cite{shor1995, steane1996, calderbank1996, laflamme1996, gottesman1997, devitt2013}, an essential tool from quantum computation to protect quantum states from errors, is a potential solution to circumvent the noise problem of quantum metrology \cite{kessler2014, dur2014, arrad2014, lu2015, unden2016, sekatski2017, demkowicz2017, zhou2018, layden2018, layden2019, ma2019, gorecki2020}. A proof of principle for quantum error correction enhanced magnetometry was outlined in \cite{herrera2015}, and further enhanced with post selection in \cite{matsuzaki2017}.

In the more recent results for quantum error correction enhanced quantum metrology, it has been shown that if the signal and noise occur in orthogonal directions \cite{demkowicz2017, zhou2018}, then error correction can mitigate the effects of noise without affecting the signal. Furthermore, it was shown that the Heisenberg limit is recoverable when the time between error correction applications can be taken as arbitrarily small. Unfortunately, this mathematical assumption is impractical with our current quantum technologies: the timescale of current error correction schemes are far from infinitesimally small, instead they are comparable to that of the dephasing times for both spin qubit systems \cite{dutt2007, taminiau2014} and superconducting qubit systems  \cite{cramer2016, ofek2016}. Moreover, a realistic error correction strategy is hindered by other factors such as noisy ancillary qubits and imperfect error correction.

In this article we consider a more pragmatic approach to incorporate error correction into quantum metrology - by accounting for the impediments one would face with current quantum technologies. We begin this article by reviewing the fundamentals of phase estimation in a noisy environment without error correction and compare it to a model enhanced with a parity check error correction strategy. We derive the necessary conditions to achieve Heisenberg-like precision, and unsurprisingly, we show that it is not permanently achievable once one discards the assumption of being able to perform arbitrarily fast error correction. Next, we derive the effects of noisy ancillary qubits and imperfect error correction has on the precision. From which, we investigate the limitations of current quantum technologies, determining which factors need to be improved upon to enable Heisenberg-level precision. Even though the results of this study are derived by examining a specific error correction strategy, we conclude by suggesting that they can be generalized to all error correcting codes.

\section{Noisy Quantum Metrology}

The canonical example of a noisy quantum metrology scheme involves $n$ qubits governed by two interactions. The first is a signal $\omega$ which causes a detuning in each of the qubits, represented by $H = \frac{\hbar \omega}{2} \sum_{m=1}^n Z_m$. The second, an interaction with the environment which causes dephasing (with rate $\gamma$ in the $X$ direction. Here $(Z_m,X_m )$ are the usual Pauli operators for the $m$th qubit. Lastly, the qubit evolves in accordance to its natural resonance frequency, which we assume is known to a high degree of precision. In the rotating reference frame, where the natural frequency of the qubit is suppressed, the Lindbladian master equation can be written as \cite{rivas2012}
\begin{equation}
\label{eq:Master}
\frac{d \rho}{dt}=-\frac{i}{\hbar} [H, \rho ] + \gamma \sum_{m=1}^n (X_m \rho X_m - \rho).
\end{equation}
The master equation can be further generalized to include a dephasing term which commuted with the signal. However, these errors cannot be efficiently corrected because one cannot differentiate between the signal and those errors \cite{demkowicz2017, zhou2018}. Hence, for off-axis Markovian noise, we can only correct its perpendicular component. In the non-Markovian case, where the noise commutes with the signal but is spatially correlated, one can implement more complicated error correction strategies \cite{layden2018, layden2019} to mitigate the effects of noise; we do not consider these noise models in this work.

The protocol of the quantum metrology scheme is to let the system evolve for time $t$, after which an appropriate measurement is performed. The likelihood of the measurement outcomes will be dependent on $\omega$, so by repeating the prepare and measure protocol many times one can precisely establish the value of $\omega$ \cite{toth2014}. As the noisy system continues to evolve it becomes significantly more difficult to distinguish the effects of the signal Hamiltonian and the effects of the noise; increasing the uncertainty of the estimate.

The quantum Fisher information (QFI) is a quantity which is inversely proportional to the minimum uncertainty of the estimate \cite{helstrom1976, braunstein1994}. If initial quantum state is an $n$ qubit Greenberger–Horne–Zeilinger (GHZ) state, then in a noiseless environment after sensing time $t$ the QFI is given by \cite{giovannetti2006}
\begin{equation}
\mathcal{Q}_\text{noiseless} = n^2 t^2.
\end{equation}
This quadratic scaling in $n$ is known as the Heisenberg limit, the ultimate precision quantum physics allows. However, in a noisy environment, the achievable precision is ultimately bounded because of said noise. It is straightforward to compute (see Appendix A) the QFI in the noisy case
\begin{equation}
\label{eq:noisyapprox}
\mathcal{Q}_\text{noisy} = n^2 t^2 \Big(1-\big(2-\frac{4}{3n} \big) \gamma t \Big) +\mathcal{O}(\gamma^2 t^4).
\end{equation}
Hence in the short time limit ($\gamma t \ll 1$) where the first two non-zero terms of the Taylor expansion dominate the behaviour of the QFI, we observe that the Heisenberg limit is lost once the quantity $2\gamma t$ becomes large. This is not a practical time scale.

\section{Error Correction Enhanced Quantum Metrology}

\begin{figure}[t]
  \centering\includegraphics[width=0.95\textwidth]{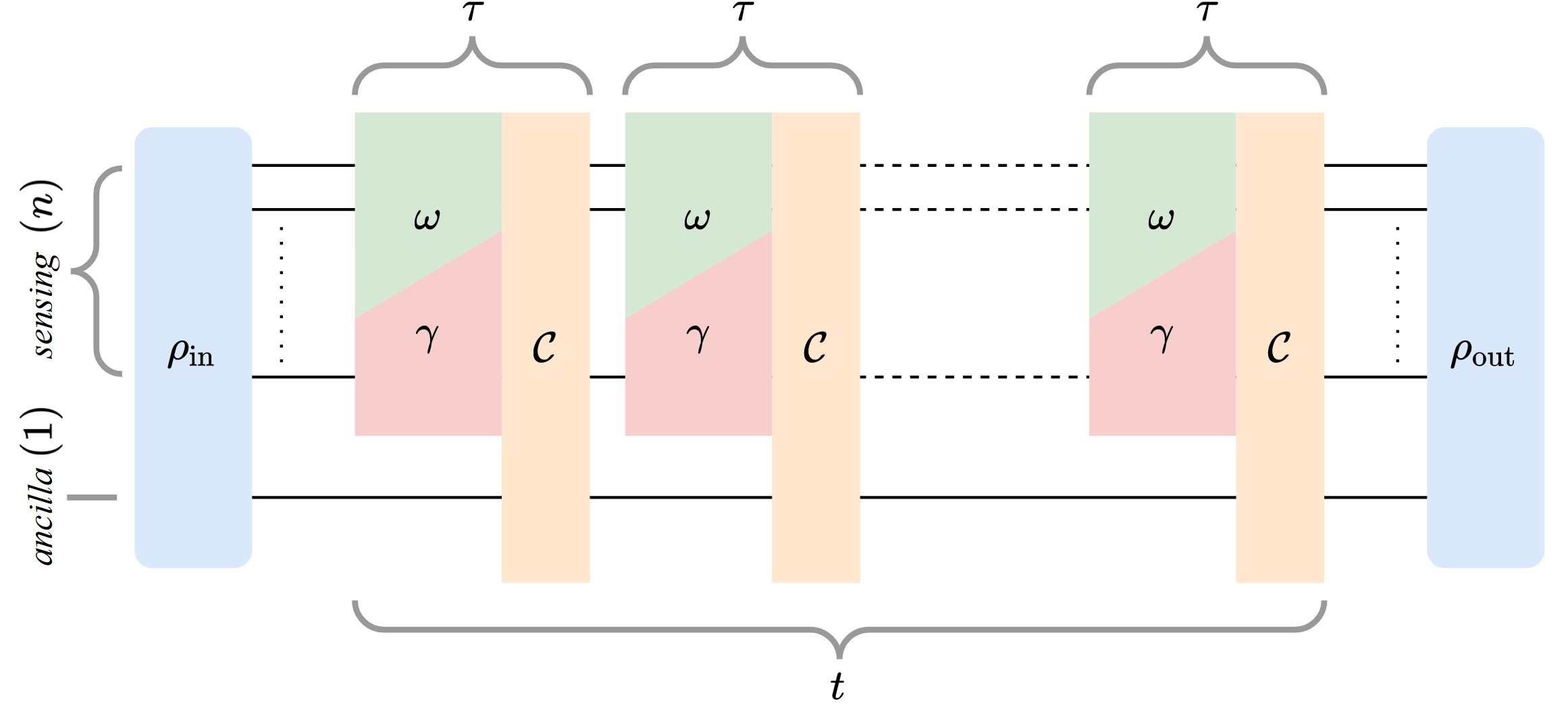}
  \captionsetup{margin=0cm}
  \caption{Schematic illustration of the error correction enhanced quantum metrology. The input state $\rho_\text{in}$, is composed of $n$ sensing qubits and one ancillary qubit. The sensing qubits evolve according to the signal $\omega$ and noise $\gamma$. The parity check code, denoted by $\mathcal{C}$, is repeatedly applied after a given time $\tau$. The final quantum state used for parameter estimation, $\rho_\text{out}$, undergoes $t/\tau$ rounds of error correction. The scheme can easily be generalized; allowing for arbitrary input states, error correction strategies and more ancillary qubits.}
  \label{fig:ECC_QM}
\end{figure}

With the issue of noise in mind, let us now incorporate error correction protocols into the quantum metrology scheme as a potential solution. Current literature suggests that noise which is perpendicular to the signal can be mitigated by performing discrete applications of error correction in rapid succession \cite{demkowicz2017, zhou2018}. However, to successfully mitigate the noise and recover the Heisenberg limit, the time between applications of error correction is taken to be arbitrarily small. Further, these studies include the assumption of perfect error correction and access to noiseless ancillary qubits. These assumptions are implausible for near term implementations of repeated error correction \cite{dutt2007, taminiau2014, cramer2016, ofek2016}.

A practical approach is to make no assumptions regarding the time it takes to perform error correction; similar to what was done in one of the original formulations of error correction and quantum metrology \cite{kessler2014}. Unfortunately, higher order error terms in \cite{kessler2014} are disregarded, which is similarly presumptuous regarding the current ability of quantum technologies. We overcome these restrictions by finding an exact solution of a specific error correction strategy: a parity check error correction code \cite{hsieh2009, fujiwara2015, roffe2018} involving a single ancillary qubit, depicted in Fig.~\ref{fig:ECC_QM}. This scheme could be implemented for instance in a hybrid quantum system involving an electron/nuclear spin system. Here the electron spin could be used a sensing qubit while the nuclear spin would act as an ancillary qubit \cite{dutt2007, taminiau2014, waldherr2014}. The parity check code is composed of two components. The first is $n$ non-destructive parity measurements between individual sensing qubits and the ancillary qubit. The second is the correction to any qubits in which an error is detected \cite{note1}.

In our model, the quantum state is initialized as an $n+1$ qubit GHZ where $n$ qubits are used for sensing while the remaining one is used as an ancilla for error correction. The sensing qubits are influenced by a signal $\omega$, as well as the noise $\gamma$. The parity check code is applied after time $\tau$ to mitigate the effects of the noise; the procedure is then repeated $t/\tau$ times where $t$ is the total sensing. Let us now determine the criteria to achieve the Heisenberg limit. Initially we  assume the ancillary qubit is noiseless and error correction is perfect, after which we consider the scenarios without these assumptions.

\subsection{Noiseless Ancilla and Perfect Error Correction}

\begin{figure*}[t]
\centering
\subfloat{\includegraphics[width=\textwidth]{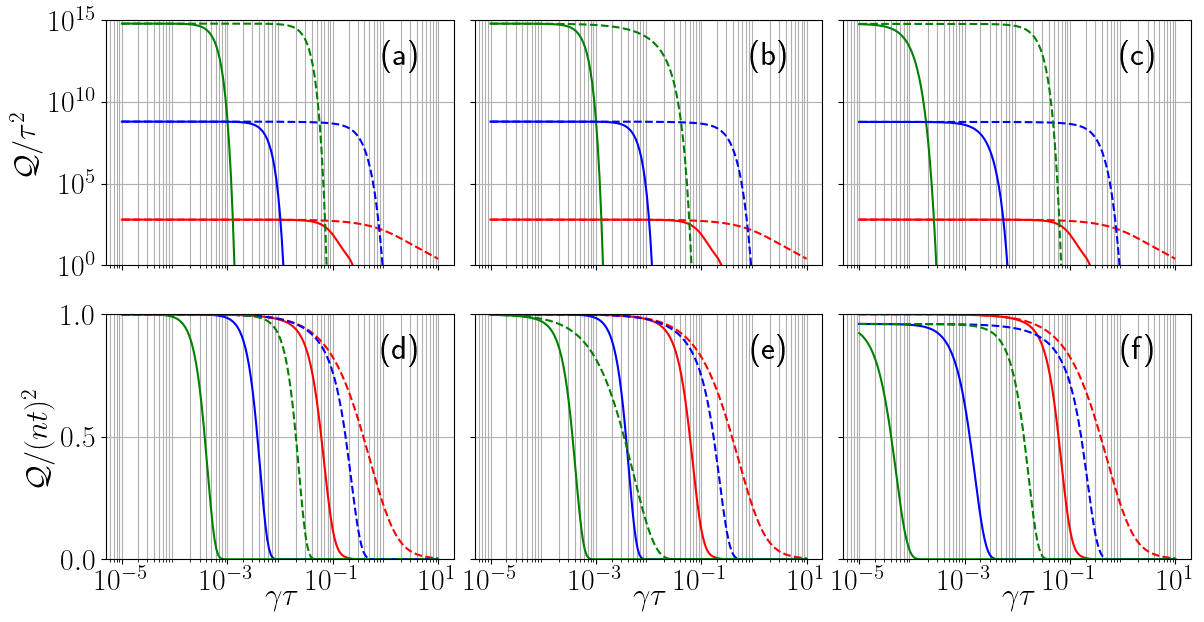}}
\vspace{5pt}
\subfloat{\includegraphics[width=0.39\textwidth]{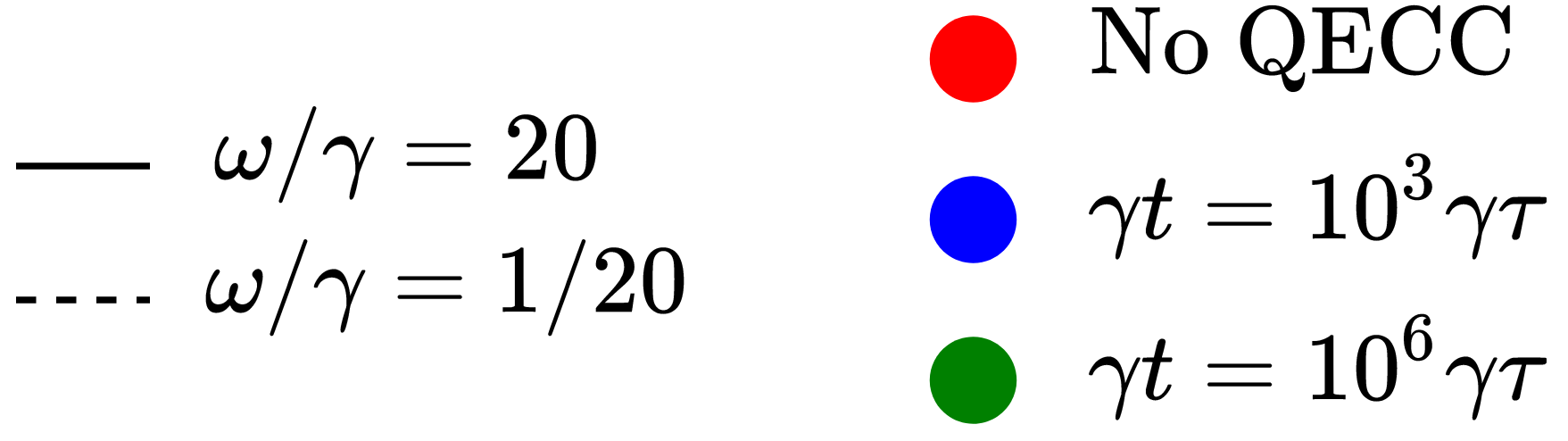}}
\vspace{3pt}
\caption{Plot of $\mathcal{Q}/\tau^2$ for an $n=25$ qubit GHZ state after undergoing repeated error correction with (a) a noiseless ancilla and perfect error correction, (b) a noisy ancilla ($\xi/\gamma=10^{-4}$) and imperfect error correction, and (c) a noiseless ancilla and imperfect error correction ($p=0.01$), with total sensing times $t/\tau=10^3,10^6$. The characteristics of a noisy state without the inclusion of a quantum error correction code (QECC) after sensing time $\tau$ is also displayed. Note that these curves are cutoff when $\mathcal{Q}/\tau^2=1$ for clarity purposes. Additionally, we illustrate the corresponding normalized QFI curves, $\mathcal{Q}/(nt)^2$, in plots (d), (e) and (f) respectfully, to emphasize the deviation from the Heisenberg limit.}
\label{fig:QFIPlots}
\end{figure*}

\begin{figure*}[t]
\centering
\subfloat{\includegraphics[width=\textwidth]{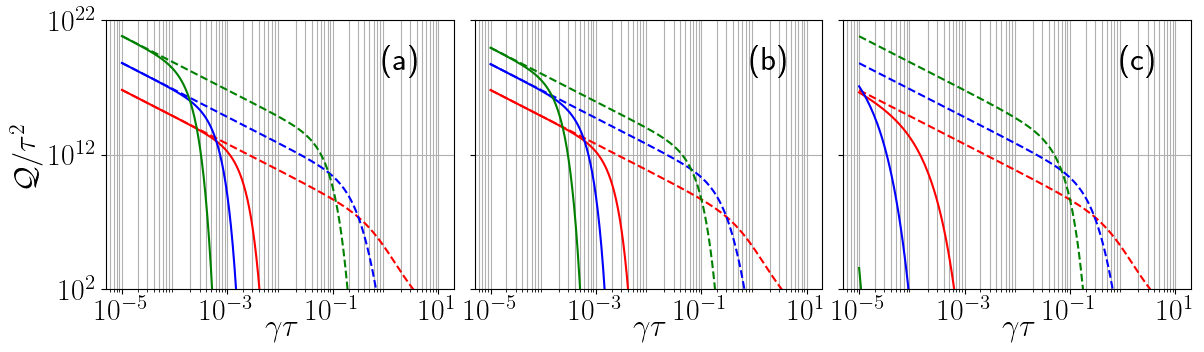}}
\vspace{5pt}
\subfloat{\includegraphics[width=0.39\textwidth]{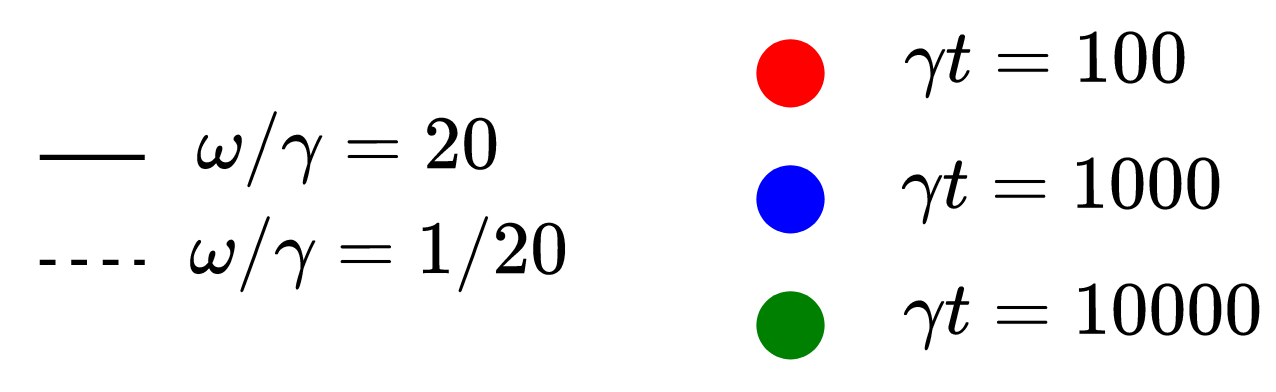}}
\vspace{3pt}
\caption{An alternative perspective on illustrating the tendencies of $\mathcal{Q}/\tau^2$ for an $n=25$ qubit GHZ state after undergoing repeated error correction. Here, the total sensing time $\gamma t$ is held constant and deviations from the linear curve on the log-log plot represent the QFI tending away from the Heisenberg limit and towards a QFI of zero. Once again we plot the three scenarios: (a) a noiseless ancilla and perfect error correction, (b) a noisy ancilla ($\xi/\gamma=10^{-4}$) and imperfect error correction, and (c) a noiseless ancilla and imperfect error correction ($p=0.01$). Note that without any error correction, the QFI after total sensing times $\gamma t=100,1000,10000$ is effectively zero.}
\label{fig:QFIPlots2}
\end{figure*}

In the ideal error correction scenario (noiseless ancilla and perfect error correction), after $t/\tau$ rounds of error correction, the quantum state can be expressed as a bipartite mixed state
\begin{equation}
\label{eq:finalstate}
\rho = \frac{1+r^{nt/\tau}}{2}\dyad*{\psi_+} + \frac{1-r^{nt/\tau}}{2} \dyad*{\psi_-},
\end{equation}
where
\begin{equation}
\ket*{\psi_\pm}= \frac{1}{\sqrt{2}} \big(\ket{0}^{\otimes n+1} \pm e^{i n \phi t/\tau} \ket{1}^{\otimes n+1} \big),
\end{equation}
and
\begin{equation}
\label{eq:small_r}
 re^{\pm i \phi}= e^{-\gamma \tau} \Big(\cos \big(\Delta \tau \big)+ \frac{\gamma \pm i \omega}{\Delta}\sin \big( \Delta \tau \big) \Big),
\end{equation}
with $\Delta=\sqrt{\omega^2-\gamma^2}$. As we are considering the quantum state immediately after the $t/\tau$th application of error correction, we of course obtain a mixture of GHZ-like states. The derivation of Eq.~(\ref{eq:finalstate}) can be found in Appendix~B. It is important to note that $r^2<1$ if $\gamma \tau > 0$. Consequently, the quantum state becomes more mixed (and less useful for quantum metrology) once the quantity $n t/\tau$ becomes very large. In Appendix B, we show that the QFI of such a quantum state can be written in the form
\begin{equation}
\label{eq:QFI_1}
\mathcal{Q}_1 = n^2 t^2 r^{2nt/\tau} f,
\end{equation}
where for small times $\tau$,
\begin{equation}
\label{eq:propconst1}
f = 1-2\gamma \tau + \mathcal{O} \big( \gamma^2 \tau ^2 \big).
\end{equation}
We immediately observe that that a Heisenberg level of precision is obtained if two conditions are met. The first condition is that $2\gamma \tau \ll 1$; it was derived in \cite{demkowicz2017, zhou2018} and is synonymous to the noisy scenario without error correction given in Eq.~(\ref{eq:noisyapprox}). The second condition is that $r^{2nt/\tau} \approx 1$, which suggests that the Heisenberg limit cannot be maintained indefinitely in a noisy environment ($r^2 < 1$) and that the QFI will eventually tend to zero. For small $\tau$ we have
\begin{equation}
r^{2nt/\tau} = 1-\frac{4}{3}n (\omega \tau)^2 \gamma t+\mathcal{O} \big( \gamma^3 \tau^3 \big),
\end{equation}
meaning we can re-write the second condition as $\frac{4}{3} n \omega^2 \tau^2 \gamma t \ll 1$. This second condition is of second order with respect to $\tau$, which is the reason it was overlooked in previous studies \cite{demkowicz2017, zhou2018}.

Both conditions are clearly illustrated in Fig.~\ref{fig:QFIPlots}a and Fig.~\ref{fig:QFIPlots}d, where we have two different family of curves plotted for a $n=25$ qubit GHZ state. The first family of curves is with $\omega/\gamma = 20$, and shows the Heisenberg limit level of precision is lost once $r^{2nt/\tau}$ begins to tend to zero. The second family of curves is for $\omega/\gamma = 1/20$, and the Heisenberg limit level of precision is lost once $\gamma \tau \approx 10^{-2}$, regardless of if $t=10^3\tau$ or $t=10^6\tau$. The two conditions are similarly illustrated in Fig.~\ref{fig:QFIPlots2}(a), where a deviation from the Heisenberg limit is again caused by the $r^{2nt/\tau} \approx 1$ condition when $\omega/\gamma=20$, and by the $2\gamma \tau \ll 1$ condition when $\omega/\gamma=1/20$. Additionally, we observes that once the QFI begins to tend away from the Heisenberg limit, the rate at which the QFI tends to zero is faster for larger values of $\gamma t$.

The reason for the stark contrast in the two families of curves ($\omega^2 \gg \gamma^2$ versus $\omega^2 \ll \gamma^2$), is due to larger deviations from the ideal case when $\omega^2 \gg \gamma^2$. Information about $\omega$ is stored in the relative phase, $n \phi t/\tau $, and if an error does occur between applications of error correction, the phase will deviate further from the ideal case. Thus, each round of error correction introduces a small amount of variance to the phase, and this variance is larger for larger values of $\omega$.

In the noisy scenario without error correction, it has been shown that the optimal sensing time (which maximizes the QFI) is $t_\text{opt} \approx 1/(n \gamma)$ \cite{chaves2013}. We can compute a similar quantity in the scenario with error correction by first noting that $\frac{\partial}{\partial_t} f = \mathcal{O} \big( (\gamma \tau )^2 \big)$. Thus, one can approximate the optimal sensing time by maximizing the quantity $t^2r^{2nt/\tau}$. In doing so one obtains
\begin{equation}
    t_\text{opt} = \frac{1}{\frac{2}{3}n \gamma \omega^2 \tau^2+\mathcal{O} \big( \gamma^2 \tau^3 \big)}.
\end{equation}
As expected, $t_\text{opt}$ increases as $\tau$ decreases, and decreases as $n$ increases. The dependence on $\omega$ can once again be linked to effective variance in the phase of the quantum state.

\subsection{Noisy Ancilla and Perfect Error Correction}

To further augment the reality of our quantum metrology scheme we impede the error correction by adding a dephasing rate, $\xi$, to the ancillary qubit. This modification results in a QFI of the form
\begin{equation}
\label{eq:noisyQFI2}
\mathcal{Q}_2 =n^2 t^2 r^{2nt/\tau}(f-g\xi) + \mathcal{O} \big( \xi^2 \tau^2 t^2 \big),
\end{equation}
where $g$ is bounded by (see Appendix B)
\begin{equation}
\Big(\frac{2}{3}-7\gamma \tau \Big)t \leq g + \mathcal{O} \big( \gamma \tau^2 \big) \leq \frac{5}{2}(t+\tau),
\end{equation}
which we can interpret as a third condition that is needed to be satisfied to obtain Heisenberg-like scaling: $\xi t \ll 1$. This is not very surprising; once $\xi t$ becomes significantly large, the ancillary qubit will be too noisy for practical error correction. This additional condition is displayed in Fig.~\ref{fig:QFIPlots}b and Fig.~\ref{fig:QFIPlots}e, where we set the ancillary qubit to have a dephasing rate of $\xi/\gamma=10^{-4}$. As expected, the noisy ancilla causes the Heisenberg limit to be lost sooner when compared to the case with a noiseless ancilla. The impact is much more prominent for the curve with $\omega/\gamma=1/20$ and $t=10^6 \tau$, where the loss of the Heisenberg limit is due to $\xi t$ becoming too large instead of $\gamma \tau$. In Fig.~\ref{fig:QFIPlots2}(b), where the total sensing time $\gamma t$ is static (and thus $\xi t$ is static), we observe a noticeable decrease in the achievable QFI when the total sensing time is largest, $\gamma t =10000$, when compared to the same set of curves in Fig.~\ref{fig:QFIPlots2}(a), where the ancillary qubit is noiseless. The additional decrease due to a noisy ancillary qubit can be overcome by occasionally re-initializing the ancillary qubit before it gets too noisy.

\subsection{Noiseless Ancilla and Imperfect Error Correction}

The second hindrance we explore is the inclusion of imperfect error correction. We simulate this by adding a probability $p$ that a parity check outputs the wrong outcome, which results in an unnecessary bit-flip correction (or lack there of). In this scenario, the QFI can be written as
\begin{equation}
\label{eq:noisyQFI3}
\mathcal{Q}_3=n^2 t^2 (rq)^{2nt/\tau}h,
\end{equation}
with
\begin{equation}
q^{2nt/\tau}=1-4p(1-p)\omega^2 t \tau  + \mathcal{O} \big( \gamma^2 \tau^2 \big),
\end{equation}
and
\begin{equation}
h=(1-2p)^2f+4p\Big( \frac{1-p}{n}+1-2p \Big) \frac{\tau}{t} + \mathcal{O} \big( \gamma^2 \tau^2 \big).
\end{equation}
Notice that the inclusion of imperfect error correction makes the Heisenberg limit unattainable; $\mathcal{Q} \rightarrow n^2 t^2 (1-2p)^2$ as $\tau \rightarrow 0$. The multiplicative factor $(1-2p)^2$ is is due to uncertainty in the error correction propagating to the uncertainty in the parameter estimation.

The additional conditions are illustrated in Fig.~\ref{fig:QFIPlots}c, Fig.~\ref{fig:QFIPlots}f and Fig.~\ref{fig:QFIPlots2}c, where we have set $p=0.01$. Firstly, as $\gamma \tau \rightarrow 0$, $\mathcal{Q}/(nt)^2 \rightarrow (1-2p)^2 \approx 0.96$. Secondly, the additional condition $q^{2nt/\tau} \approx 1$ is again associated with additional variance being added to the relative phase because of the imperfect error correction. The impact of which is much more influential when $\omega/\gamma=20$, as opposed to when $\omega/\gamma = 1/20$.

\subsection{Limitations of Current Quantum Technologies}

\begin{figure}[t]
\centering
\begin{minipage}[c]{.73\textwidth}
\centering
\subfloat{\includegraphics[width=\textwidth]{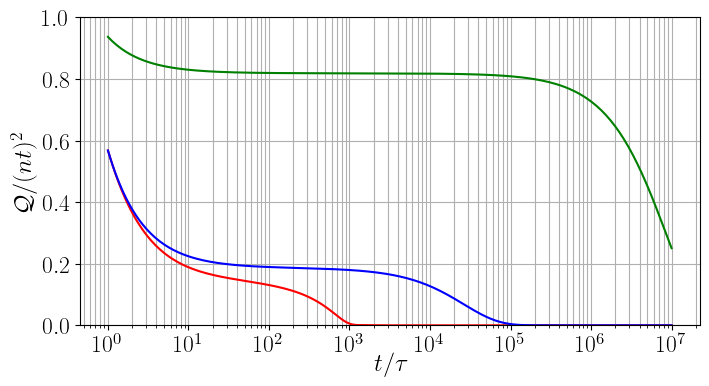}}
\end{minipage}%
\hspace{0.03\textwidth}
\begin{minipage}[c]{.15\textwidth}
\centering
\subfloat{\includegraphics[width=0.97\textwidth]{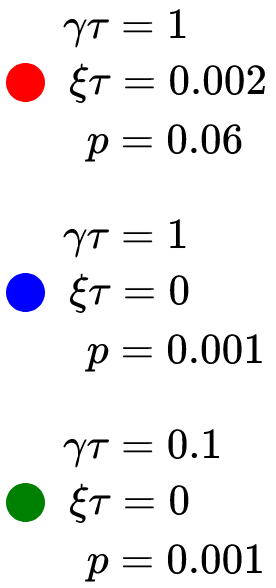}}
\vspace{18pt}
\end{minipage}
\caption{Normalized QFI in the small signal regime $\omega/\gamma =0.01$. Using values from today's quantum technologies ($\gamma^{-1} = \tau = 10^{-6}$s, $\xi^{-1}= 5\times10^{-4}$s, and $p=0.06$) \cite{dutt2007, taminiau2014} (\tikzcircle{2.8pt}) a QFI of $\sim 20\%$ of the Heisenberg limit can be achieved for sensing times $t=10^1 \tau$. With improved error correction fidelity and a noiseless ancilla (\tikzcircle[blue,fill=blue]{2.8pt}) this can be sustained for a sensing time $t=10^3 \tau$. The QFI is significantly improved when $\gamma \tau=0.1$ (\tikzcircle[green2,fill=green2]{2.8pt}).}
\label{fig:QFIBenchmarking}
\end{figure}

In \cite{dutt2007, taminiau2014}, the electron spin of a nitrogen-vacancy center is entangled to carbon-13 nuclear spins, the nuclear spins act as ancillary qubits, and error correction is performed on the electorn spin. The reported dephasing rates are $\gamma^{-1} \sim 10^{-6}$s and $\xi^{-1} \sim 5\times10^{-4}$s. The error correction is being performed on a similar timescale of $\tau \sim 10^{-6}$s, with infidelity reported at $p=0.06$ \cite{taminiau2014}. In Fig.~\ref{fig:QFIBenchmarking}, we benchmark the sensing capability of a single qubit systems when $\omega^2 \ll \gamma^2$, using the listed values. As expected, the Heisenberg limit is unattainable. Moreover, it is still unattainable if one performs near perfect error correction ($p=0.001$) and uses noiseless ancillary qubits ($\xi=0$) which can be approximately achieved by frequently re-initializing the ancilla before it becomes too noisy. Notably though, when the quantity $\gamma \tau$ decreases by a factor of ten, one attains a QFI of $\sim 80\%$ of the Heisenberg limit for a total sensing time $t=10^5 \tau$; greatly outclassing the precision achieved in current experiments \cite{taylor2008, wasilewski2010, razzoli2019} using spin systems. Of course, this result should be used with caution. A realistic metrology scheme is hindered by other factors not discussed in this study, such as imperfect resources and noise parallel to the signal, which cannot be suppressed with error correction.

\section{Other Error Correction Strategies}

Until now, we have only discussed one possible error correction protocol, whereas \cite{demkowicz2017, zhou2018} make no assumptions regarding the error correction strategy. Although a completely general result is more satisfying, it is unfeasible with our solution method. However, using our mathematical methodology, one can substitute any specific error correction strategy in place of the parity check error correction code. Alternatively, one could consider a continuous error correction strategy \cite{paz1998, ahn2002, sarovar2005} by appropriately modifying the master equation, Eq.~(\ref{eq:Master}). Note that continuous error correcting codes are difficult to implement \cite{sarovar2005} and were thus not considered in this work. Finally, one could forego error correction altogether and adapt our model to one of dynamical decoupling \cite{viola1999, sekatski2016} or reservoir engineering \cite{schirmer2010}.

We conjecture that, just as with discrete applications of the parity check code, for any noise mitigation strategy, the QFI will depend on term similar to $r^{2nt/\tau}$: one which suggests that there exist a critical time where the QFI begins to tend to zero. In fact, we show in Appendix C that using the $n$ qubit bit flip code \cite{gottesman1997} yields the results
\begin{equation}
\mathcal{Q} = n^2 t^2 r^{2nt/\tau}f+\mathcal{O} \big( (\gamma \tau)^{\frac{n-1}{2}} \big).
\end{equation}
Hence, for large $n$ the QFI using the bit flip code is effectively the same as if one utilizes the parity check code. The reasoning supporting the aforementioned conjecture is that any errors which occur will cause the relative phase to deviate from the ideal value, and the deviation will remain even after performing a correction. Thus, after each round of error correction, the variance in the phase will increase, which propagates to an increase in variance in the eventual estimate of $\omega$. This conjecture is easily extended to any realistic noise model; as one expects the relative phase to deviate from the ideal value after performing error correction, regardless of the noise model.

We are not suggesting that the parity check code is the most efficient noise mitigation strategy, for orthogonal noise, at retaining the Heisenberg limit. For example, if one performs adaptive parameter estimation \cite{gill2005, fujiwara2006}, one could supplement the error correction strategy with an inverse Hamiltonian to correct some of the phase deviations. This strategy is more difficult to implement, but most likely outperforms the parity check code. However, it would still fail to perfectly correct all of the deviations in the phase. To do so, and hence maintain the Heisenberg limit indefinitely, would require knowledge of the exact time errors occur at, as well as exact knowledge of $\omega$ (note that the latter requirement defeats the purpose of quantum metrology). Of course, as quantum technologies continue to improve, and the frequency at which these noise mitigation tools can be applied increases, so too does our ability to maintain the Heisenberg limit for increased sensing times.

\section{Remarks}

In this article we analysed the ability of an error correction protocol to recover the Heisenberg limit in noisy quantum metrology. As expected from previous results \cite{demkowicz2017, zhou2018}, the Heisenberg limit can be achieved for longer duration's when compared to a similar scheme without error correction and is conditioned on the fact that $2\gamma \tau \ll 1$. Notably though, our results suggest that the Heisenberg limit is not permanently achievable, and is lost once the quantity $n t/\tau$ becomes too large. This previously undiscussed requirement is due to the fact that a small amount of variance is introduced to the relative phase after each round of error correction; eventually the total variance is too large to overcome. We also showed that any uncertainty in the error correction will propagate to the QFI, making the actual Heisenberg limit unachievable, regardless of the speed of the frequency of error correction. Moreover, the uncertainty of error correction increases the added variance to the relative phase.

A more practical figure of merit for quantum metrology is the Fisher information \cite{braunstein1994} (instead of quantum Fisher information). However, the Fisher information is dependent on a specific measurement strategy, whereas the QFI is optimized over all measurement strategies. Thus, to obtain analytic results that are not strategy dependent, we compute the QFI. Nonetheless, we provide sufficient mathematical results in the Appendix to compute the Fisher information for a given measurement strategy. Furthermore, we show in Appendix~D that there exists a measurement strategy using single qubit measurements such that the Fisher information is approximately equal to the QFI for the error correction enhanced quantum metrology scheme.

We chose to specifically analyse the effects of repeated error correction in the scenario when the quantum state used for sensing is initialized in an $n$ qubit GHZ state. The logical generalization would be to expand the results to a broader scope of initial states; such as squeezed states \cite{gross2012}, symmetric states \cite{oszmaniec2016}, or graph states \cite{shettell2020}. It is possible that these quantum states (which do not achieve the Heisenberg limit, but do achieve a quantum advantage) are more robust to the effects of noise and can maintain a quantum advantage for a longer total sensing time.

\section*{Acknowledgements}

Nathan Shettell and Damian Markham acknowledge support of the ANR through the ANR-17-CE24-0035 VanQuTe. Nathan Shettell would also like to acknowledge the support of the NII international internship program.

\appendix

\section{The QFI of a Noisy GHZ without Error Correction}

We wish to solve the matrix equation
\begin{equation}
\label{Eq:MEqnApp}
\frac{d \rho}{dt}=-\frac{i}{\hbar} [H, \rho ] + \gamma \sum_{m=1}^n (X_m \rho X_m - \rho),
\end{equation}
with $H = \frac{\hbar \omega}{2} \sum_{m=1}^n Z_m$. Without loss of generality, we can assume that the solution is of the form
\begin{equation}
\rho = \sum_{j,k} \alpha_{j,k} \dyad{j}{k},
\end{equation}
where $j,k \in \{0,1 \}^{\otimes n}$ are bit strings of length $n$. To obtain a solution, we instead consider Eq.~(\ref{Eq:MEqnApp}) as a set of linear differential equations and solve for the amplitudes $\alpha_{j,k}$. Since we are considering the scenario where the quantum state is initialized in a GHZ state, the only non-zero of the amplitudes are those of the form $\alpha_{j,j}$ or $\alpha_{j,\bar{j}}$ (where $\ket{\bar{j}}=X^{\otimes n}\ket{j}$.

We set $\vec{a}$ to be the vector of size $2^n$ containing all amplitudes of the form $\alpha_{j,j}$. Similarly, we set $\vec{b}$ to be the vector of size $2^n$ containing all amplitudes of the form $\alpha_{j,\bar{j}}$. Both vectors are ordered with respect to ascending values of $j$. This framework transforms Eq.~(\ref{Eq:MEqnApp}) into two separate matrix differential equations, i) $\frac{d \vec{a}}{dt} = A \vec{a}$, and ii) $\frac{d \vec{b}}{dt} = B \vec{b}$, where
\begin{align}
A &= \sum_{m=0}^{n-1} \begin{pmatrix}
1 & 0 \\
0 & 1
\end{pmatrix}^{\otimes m} \otimes \begin{pmatrix}
-\gamma & \gamma \\
\gamma & -\gamma
\end{pmatrix} \otimes \begin{pmatrix}
1 & 0 \\
0 & 1
\end{pmatrix}^{\otimes n-m-1}, \\
B &= \sum_{m=0}^{n-1} \begin{pmatrix}
1 & 0 \\
0 & 1
\end{pmatrix}^{\otimes m} \otimes \begin{pmatrix}
-i \omega -\gamma & \gamma \\
\gamma & i \omega -\gamma
\end{pmatrix} \otimes \begin{pmatrix}
1 & 0 \\
0 & 1
\end{pmatrix}^{\otimes n-m-1}.
\end{align}
The solutions of the differential matrix equations are
\begin{equation}
\label{eqn:Aevo}
\vec{a} = e^{A t} \vec{a}_0 = e^{-n \gamma t} \begin{pmatrix}
\cosh ( \gamma t ) & \sinh ( \gamma t ) \\
\sinh ( \gamma t ) & \cosh ( \gamma t )
\end{pmatrix}^{\otimes n} \vec{a}_0,
\end{equation}
and
\begin{equation}
\label{eq:Bevo}
\begin{aligned}
\vec{b} = e^{B t} \vec{b}_0 &= e^{-n \gamma t} \begin{pmatrix}
\cos ( \Delta t ) - i \frac{\omega}{\Delta} \sin ( \Delta t ) & \frac{\gamma}{\Delta} \sin ( \Delta t ) \\[5pt]
\frac{\gamma}{\Delta} \sin (\Delta t ) & \cos ( \Delta t ) + i \frac{\omega}{\Delta} \sin ( \Delta t )
\end{pmatrix}^{\otimes n} \vec{b}_0 \\
&= e^{-n \gamma t} \begin{pmatrix}
x_- & y \\
y & x_+
\end{pmatrix}^{\otimes n} \vec{b}_0,
\end{aligned}
\end{equation}
where $\vec{a}_0$ and $\vec{b}_0$ are the initial amplitude vectors and $\Delta=\sqrt{\omega^2 - \gamma^2}$. The symbolic representation of the entries of the matrix $B$ ($x_\pm$ and $y$) is to add clarity in latter computations.

The final quantum state can be expressed in the form
\begin{equation}
\label{Eq:GenSoln}
\rho = \frac{1}{2} \sum_j \lambda_{j,+} \dyad*{\psi_{j,+}}+\lambda_{j,-} \dyad*{\psi_{j,-}},
\end{equation}
with
\begin{equation}
\lambda_{j,\pm}=e^{-n \gamma t} \frac{s_j \pm r_j}{2},
\end{equation}
and
\begin{equation}
\ket{\psi_{j,\pm}}=\frac{1}{\sqrt{2}}\big(e^{-i \theta_j /2}\ket{j} \pm e^{+i \theta_j /2} \ket{\bar{j}} \big). 
\end{equation}
The factor of $1/2$ in front of the sum is to avoid double counting, because $\lambda_{j,\pm}=\lambda_{\bar{j},\pm}$ and $\ket{\psi_{j,\pm}}=\ket{\psi_{\bar{j},\pm}}$. The eigenvalues and eigenvectors are parameterized by
\begin{align}
s_j &= \cosh^{n-h_j}(\gamma t) \sinh^{h_j}(\gamma t)+\cosh^{h_j}(\gamma t) \sinh^{n-h_j}(\gamma t), \\[5pt]
r_j e^{\pm i \theta_j} &= x_\pm^{h_j} y^{n-h_j}+x_\mp^{n-h_j} y^{h_j},
\end{align}
where $h_j$ is the Hamming weight (number of $1$'s) of $j$.

The QFI of a quantum state $\sigma=\sum_{j} \lambda_j \dyad{\psi_j}$, where $\{ \ket{\psi_j} \}$ form an orthonormal basis and $\sum_j \lambda_j=1$, is \cite{braunstein1994}
\begin{equation}
\mathcal{Q}=\sum_{\substack{ j \\ \lambda_j \neq 0}} \frac{\dot{\lambda}_j^2}{\lambda_j}+2\hspace{-7pt}\sum_{\substack{ j,k \\ \lambda_j + \lambda_k \neq 0}}\hspace{-7pt} \frac{(\lambda_j - \lambda_k)^2}{\lambda_j+\lambda_k} \big| \braket*{\dot{\psi}_j}{\psi_k} \big|^2,
\end{equation}
where we have used the notation $\dot{\square}=\partial_\omega \square$ for clarity. Combining this with the final quantum state $\rho$, one can obtain the QFI of a noisy GHZ without error correction. We again float a factor of $1/2$ in front of the sum to avoid double counting. In doing so one obtains
\begin{equation}
\begin{aligned}
\mathcal{Q}_\text{noisy} \hspace{-2pt} &= \hspace{-1pt}\frac{1}{2}\sum_j \Bigg( \frac{\dot{\lambda}_{j,+}^2}{\lambda_{j,+}} \hspace{-2pt} + \hspace{-2pt} \frac{\dot{\lambda}_{j,-}^2}{\lambda_{j,-}}\hspace{-2pt}+2\frac{(\lambda_{j,+}-\lambda_{j,-})^2}{\lambda_{j,+}+\lambda_{j,-}} \Big( \big| \braket*{\psi_{j,+}}{\dot{\psi}_{j,-}} \big|^2\hspace{-3pt}+\hspace{-1pt}\big| \braket*{\psi_{j,-}}{\dot{\psi}_{j,+}} \big|^2 \Big) \Bigg)\\
&= \frac{e^{-n \gamma t}}{2}\sum_j \frac{s_j \dot{r}_j^2}{s_j^2-r_j^2} +\frac{r_j^2}{s_j} \dot{\theta}_j^2 \\
&= n^2t^2\Big(1-\big(2-\frac{4}{3n}\big)\gamma t\Big)+ \mathcal{O}\big(\gamma^2 t^4 \big).
\end{aligned}
\end{equation}

\section{The QFI of a Noisy GHZ Enhanced with Repeated Error Correction}

Recall in the scenario enhanced with error correction, we let the system evolve for a total time $\tau$, after which error correction is performed. This processes is repeated until the system has evolved for total time $t$ (without loss of generality we assume that $t/\tau$ is a positive integer). Additionally, an ancillary qubit is included into the dynamics, which we set to have an index $m=n+1$.

We begin by solving a completely general scenario in which the ancilla is noisy and the error correction is imperfect, from which we can obtain varying results for all the combinations of noiseless/noisy ancilla and perfect/imperfect error correction by taking the appropriate limits.

We model the noisy ancilla by subjecting it to dephasing in the $X$ direction with rate $\xi$, this changes the matrix equations to be
\begin{equation}
e^{A \tau} = e^{-(n \gamma+\xi) \tau} \begin{pmatrix}
c_\gamma & s_\gamma \\
s_\gamma & c_\gamma
\end{pmatrix}^{\otimes n} \otimes \begin{pmatrix}
c_\xi & s_\xi \\
s_\xi & c_\xi
\end{pmatrix},
\end{equation}
and
\begin{equation}
e^{B \tau} = e^{-(n \gamma+\xi) \tau} \begin{pmatrix}
x_- & y \\
y & x_+
\end{pmatrix}^{\otimes n} \otimes \begin{pmatrix}
c_\xi & s_\xi \\
s_\xi & c_\xi
\end{pmatrix}.
\end{equation}
Here we have recycled the same notation which we used in Appendix A. To be concise we set $c_\gamma=\cosh ( \gamma \tau)$, $s_\gamma=\sinh ( \gamma \tau)$, $c_\xi=\cosh ( \xi \tau)$ and $s_\xi=\sinh ( \xi \tau)$.

Recall that in the parity check code, a correction is made on a sensing qubit if it has a different parity to the ancillary qubit. We simulate imperfect error correction by adding a probability that the syndrome measurement outputs an incorrect result with probability $p$, which would result in the incorrect correction being applied. This operation can be represented as a the matrix
\begin{equation}
E =\begin{pmatrix}
1-p & 1-p \\
p & p \\
\end{pmatrix}^{\otimes n} \otimes \begin{pmatrix}
1 & 0 \\
0 & 0 \\
\end{pmatrix} + \begin{pmatrix}
p & p \\
1-p & 1-p \\
\end{pmatrix}^{\otimes n} \otimes \begin{pmatrix}
0 & 0 \\
0 & 1 \\
\end{pmatrix}.
\end{equation}

Combining the evolution matrices, $e^{A\tau}$ and $e^{B\tau}$, and the error correction matrix, $E$, we obtain an expression for the amplitudes of the final quantum state
\begin{equation}
\label{Eq:AEvo}
\begin{aligned}
\vec{a} = \big( E e^{A\tau} \big)^{t/\tau} \vec{a}_0
=e^{-\xi t} \Bigg( & \begin{pmatrix}
1-p & 1-p \\
p & p
\end{pmatrix}^{\otimes n} \otimes \begin{pmatrix}
c_\xi & s_\xi \\
0 & 0
\end{pmatrix} \\
+& \begin{pmatrix}
p & p \\
1-p & 1-p
\end{pmatrix}^{\otimes n} \otimes \begin{pmatrix}
0 & 0 \\
s_\xi & c_\xi
\end{pmatrix} \Bigg)^{t/\tau} \vec{a}_0,
\end{aligned}
\end{equation}
and
\begin{equation}
\label{Eq:BEvo}
\begin{aligned}
\vec{b} = \big( E e^{B\tau} \big)^{t/\tau} \vec{b}_0
=r^{nt/\tau}e^{-\xi t} \Bigg( & \begin{pmatrix}
(1-p)e^{-i\phi} & (1-p)e^{i\phi} \\
pe^{-i\phi} & pe^{i\phi}
\end{pmatrix}^{\otimes n} \otimes \begin{pmatrix}
c_\xi & s_\xi \\
0 & 0
\end{pmatrix} \\
+& \begin{pmatrix}
pe^{-i\phi} & pe^{i\phi} \\
(1-p)e^{-i\phi} & (1-p)e^{i\phi}
\end{pmatrix}^{\otimes n} \otimes \begin{pmatrix}
0 & 0 \\
s_\xi & c_\xi
\end{pmatrix} \Bigg)^{t/\tau} \vec{b}_0,
\end{aligned}
\end{equation}
where we define
\begin{equation}
    re^{\pm i \phi} = e^{-\gamma \tau} ( x_\pm + y ) = e^{-\gamma \tau} \Big( \cos ( \Delta \tau ) + \frac{\gamma \pm i \omega}{\Delta} \sin ( \Delta \tau) \Big).
\end{equation}
It easy to show that from Eq.~(\ref{Eq:AEvo}) that the final amplitude corresponding to the outer product $\dyad{j0}$ is equal to $\frac{(1-p)^{n-h_j}p^{h_j}}{2}$, and similarly the amplitude corresponding to the outer product $\dyad{\bar{j}1}$ is also equal to $\frac{(1-p)^{n-h_j}p^{h_j}}{2}$. Recall that we have set the final qubit to be the ancillary qubit, and the first $n$ to be the sensing qubits. Therefore, in this scenario $h_j$ is the Hamming weight of the bit string of the sensing qubits. The solution to Eq.~(\ref{Eq:BEvo}) is more complex. However, after the first round of error-correction (and each subsequent round), the amplitude corresponding to the outer product $\dyad{j0}{\bar{j}1}$ is of the form $\frac{(1-p)^{n-h_j}p^{h_j} Re^{-i \theta}}{2}$, and the amplitude corresponding to the outer product $\dyad{\bar{j}1}{j0}$ is of the form $\frac{(1-p)^{n-h_j}p^{h_j} Re^{i \theta}}{2}$. Therefore, one can simplify the problem of solving for the final values of $Re^{\pm i \theta}$. This can be done by constructing a recurrence relation from Eq.~(\ref{Eq:BEvo}). By setting $N=n(t/\tau-1)$, we obtain
\begin{equation}
\begin{aligned}
\begin{pmatrix}
Re^{-i \theta} \\
Re^{+i \theta}
\end{pmatrix} \hspace{-3pt} &= \hspace{-1pt} r^{nt/\tau} e^{-\xi t}  \begin{pmatrix}
c_\xi q_- & s_\xi q_+ \\
s_\xi q_- & c_\xi q_+ \\
\end{pmatrix}^N \begin{pmatrix}
\upsilon_- \\
\upsilon_+
\end{pmatrix} \\
&= \hspace{-1pt} r^{n t/\tau} e^{-\xi t}  \left( \hspace{-2.5pt} \frac{\mu_+ \mu_-^N-\mu_-\mu_+^N}{\mu_+-\mu_-} \begin{pmatrix}
1 & 0 \\
0 & 1
\end{pmatrix} \hspace{-4pt}+\hspace{-1pt} \frac{ \mu_+^N-\mu_-^N}{\mu_+-\mu_-} \begin{pmatrix}
c_\xi q_- & s_\xi q_+ \\
s_\xi q_- & c_\xi q_+ \\
\end{pmatrix} \hspace{-3.5pt} \right) \hspace{-3pt} \begin{pmatrix}
\upsilon_- \\
\upsilon_+
\end{pmatrix},
\end{aligned}
\end{equation}
with
\begin{equation}
    q_\pm=(1-p)e^{\pm i \phi}+p e^{\mp i \phi},
\end{equation}
\begin{equation}
\mu_\pm = c_\xi \cos \phi \pm \sqrt{q_+ q_- s_\xi^2-(1-2p)^2 c_\xi^2 \sin^2 \phi},
\end{equation}
and 
\begin{equation}
\upsilon_\pm = c_\xi e^{\pm i n \phi} + s_\xi e^{\mp i n \phi}. 
\end{equation}

Therefore, in the general case with a noisy ancilla and imperfect error correction, the quantum state after $t/\tau$ rounds of error correction can be written as
\begin{equation}
\rho = \sum_j \lambda_{j,+} \dyad{\psi_{j,+}}+\lambda_{j,-} \dyad{\psi_{j,-}},
\end{equation}
where
\begin{equation}
\lambda_{j,\pm}=(1-p)^{n-h_j}p^{h_j}\frac{1\pm R}{2},
\end{equation}
and
\begin{equation}
\ket{\psi_{j,\pm}}=\frac{1}{\sqrt{2}} \big( e^{-i\theta/2} \ket{j0}\pm e^{+i\theta/2} \ket{\bar{j}1} \big).
\end{equation}
From which, we compute the QFI to be
\begin{equation}
\begin{aligned}
\mathcal{Q} &= \sum_j  \Bigg( \frac{\dot{\lambda}_{j,+}^2}{\lambda_{j,+}} +  \frac{\dot{\lambda}_{j,-}^2}{\lambda_{j,-}}+2\frac{(\lambda_{j,+}-\lambda_{j,-})^2}{\lambda_{j,+}+\lambda_{j,-}} \Big( \big| \braket*{\psi_{j,+}}{\dot{\psi}_{j,-}} \big|^2+\big| \braket*{\psi_{j,-}}{\dot{\psi}_{j,+}} \big|^2 \Big) \Bigg)\\
&= \frac{\dot{R}^2}{1-R^2}+R^2 \dot{\theta}^2 .
\end{aligned}
\end{equation}

We can now determine the properties of the QFI in the varying scenarios.

\begin{center}
    \textit{Case 1: Noiseless Ancilla and Perfect Error Correction}
\end{center}

In the simplest case with a noiseless ancilla ($\xi=0$) and perfect error correction ($p=0$), we obtain that $Re^{\pm i \theta}=\big( re^{\pm i \phi}\big)^{nt/\tau}$. Therefore, we can write that

\begin{equation}
    \mathcal{Q}_1 = n^2 t^2 r^{2nt/\tau} f,
\end{equation}
where
\begin{equation}
\label{eq:propconstant1}
f=\frac{1}{\tau^2}\Big( \frac{1}{1-r^{2nt/\tau}}\frac{\dot{r}^2}{r^2}+\dot{\phi}^2 \Big) = 1-2 \gamma \tau + \frac{7 \gamma^2 \tau^2}{3} + \frac{4\gamma \tau^2}{3nt} + \mathcal{O} \big(\gamma^3 \tau^3 \big),
\end{equation}
and
\begin{equation}
r^{2nt/\tau}=1-\frac{4}{3} n t \gamma \omega^2 \tau^2 + \mathcal{O} \big( \gamma^3 \tau^3 \big).
\end{equation}

It is easy to verify that $r^2 <1$ when $\gamma \tau >0$ (assuming $\omega \neq 0$), implying that the QFI will eventually tend to zero in a noisy environment. Note that we can write
\begin{equation}
r^2 = e^{-2 \gamma \tau} \Big( 1+\frac{\gamma}{\Delta} \sin \big(2 \Delta \tau \big) + \frac{2\gamma^2}{\Delta^2} \sin^2 \big(\Delta \tau \big) \Big).
\end{equation}
Suppose that $\omega^2 - \gamma^2 \geq 0$, then we have that $\Delta=|\Delta|$ is real, as well as $\sin (\Delta \tau) \leq \Delta \tau$ and $\sin (2\Delta \tau) \leq 2\Delta \tau$. Thus we have the inequality
\begin{equation}
r^2  \leq e^{-2 \gamma \tau} \Big( 1+2\gamma \tau + 2\gamma^2\tau^2 \Big),
\end{equation}
where on the right hand side of the above inequality is strictly decreasing (with respect to $\gamma \tau$) and equal to one when $\gamma \tau =0$. The other possible case is that $\omega^2-\gamma^2 <0$, then we have that $\Delta=i|\Delta|$, as well as 
\begin{equation}
\frac{\sinh \big(|\Delta| \tau)}{|\Delta|} < \frac{\sinh \big( \gamma \tau \big)}{\gamma}, \;\; \text{and} \;\; \frac{\sinh \big( 2 |\Delta | \tau \big)}{|\Delta|} < \frac{\sinh \big(2 \gamma \tau \big)}{\gamma}.
\end{equation}
From which we obtain a second inequality
\begin{equation}
r^2  < e^{-2 \gamma \tau} \Big( 1+2 \sinh (\gamma \tau) + 2\sinh^2(\gamma \tau) \Big) =1.
\end{equation}
Combining both results, assuming $\omega \neq 0$, we obtain that $r^2 < 1$ if $\gamma \tau >0$.

\begin{center}
    \textit{Case 2: Noisy Ancilla and Perfect Error Correction}
\end{center}

In the second case we have a noisy ancilla ($\xi \neq 0$) and perfect error correction ($p=0$). The analytic expression for $Re^{\pm i \theta}$ is quite complicated, so instead we expand with respect to $\xi$ to obtain

\begin{equation}
Re^{\pm i \theta} = (1-\xi t)\big( r e^{\pm i \phi} \big)^{nt/\tau} + \xi \tau \frac{\sin ( n t \phi/\tau )}{\sin ( n \phi )} \big( r e^{\mp i \phi} \big)^{n} + \mathcal{O} \big( \xi^2 \tau^2 \big).
\end{equation}

After expanding the QFI with respect to $\xi$, we obtain
\begin{equation}
\mathcal{Q}_2 =n^2 t^2 r^{2n t /\tau}(f-g \xi)+\mathcal{O} \big( \xi^2 \tau^2 t^2 \big),
\end{equation}
where
\begin{equation}
\begin{aligned}
g \hspace{-2.5pt} = &\Big(\frac{n \omega t \big(1+3 \cos (2n \omega t)\big)+(n^2 \omega^2 t^2 -2) \sin (2 n \omega t)}{n^3 \omega^3 t^3}+2 \Big)t \\
&\hspace{-2.5pt} + \hspace{-2.5pt} \frac{2\big(n \omega t \cos ( n \omega t )- \sin (n \omega t) \big)^2}{n^2 \omega^2 t^2} \tau  \\
& \hspace{-2.5pt}+\hspace{-2.5pt} \Big( \frac{(4n \omega t-2n^3 \omega^3 t^3)\cos (2n \omega t) \hspace{-2pt} - \hspace{-2pt} (2-5n^2 \omega^2 t^2) \sin (2n \omega t)}{n^3 \omega^3 t^3} \hspace{-1.5pt} - \hspace{-1.5pt} 4 \Big) \gamma t \tau \hspace{-2pt} + \hspace{-2pt} \mathcal{O} \big( \gamma \tau^2 \big).
\end{aligned}
\end{equation}
To simplify analysis, we utilize the following inequalities:
\begin{gather}
\frac{2}{3} \leq \frac{n \omega t \big(1+3 \cos (2n \omega t)\big)+(n^2 \omega^2 t^2 -2) \sin (2 n \omega t)}{n^3 \omega^3 t^3}+2  \leq \frac{5}{2}, \\
0 \leq \frac{2\big(n \omega t \cos ( n \omega t )- \sin (n \omega t) \big)^2}{n^2 \omega^2 t^2} \leq \frac{5}{2}, \\
-7 \leq \frac{(4n \omega t-2n^3 \omega^3 t^3)\cos (2n \omega t)-(2-5n^2 \omega^2 t^2) \sin (2n \omega t)}{n^3 \omega^3 t^3} -4 \leq 0,
\end{gather}
from which it follows that
\begin{equation}
\Big(\frac{2}{3}-7\gamma \tau \Big)t \leq g + \mathcal{O} \big( \gamma \tau^2 \big) \leq \frac{5}{2}(t+\tau).
\end{equation}

\begin{center}
    \textit{Case 3: Noiseless Ancilla and Imperfect Error Correction}
\end{center}

In the third case we have a noiseless ancilla ($\xi = 0$) and imperfect error correction ($p \neq 0$). It is straightforward to obtain
\begin{equation}
    Re^{\pm i \theta} = \big( r e ^{\pm i \phi} \big)^{nt/\tau} \big( q_\pm e^{\mp i \phi} \big)^{n(t/\tau-1)}.
\end{equation}

We compute the QFI to be of the form
\begin{equation}
\mathcal{Q}_3 = n^2 t^2 (rq)^{2nt/\tau} h,
\end{equation}
where we define $q^2=q_+ q_-$, which when raised to the exponent $nt/\tau$ satisfies
\begin{equation}
q^{2nt/\tau}=1-4p(1-p) \omega^2 t \tau+ \mathcal{O} \big( \gamma^2 \tau^2 \big), 
\end{equation}
and
\begin{equation}
h =(1-2p)^2 f + 4p\Big( \frac{1-p}{n}+1-2p\Big) \frac{\tau}{t} + \mathcal{O} \big( \gamma^2 \tau^2 \big).
\end{equation}

\section{The QFI when using the Bit Flip Code}

Just as in Appendix A and B we solve for the final quantum state via matrix equations. Note that the $n$ qubit bit flip code \cite{gottesman1997} does not have an ancilla, thus the whole system is $n$ qubits. The bit flip code maps the outer product $\dyad{j}{j}$ to $\dyad{0}^{\otimes n}$ if $h_j < n/2$ and $\dyad{1}^{\otimes n}$ if $h_j > n/2$ (we assume that $n$ is odd to avoid complications when $h_j = n/2$). Similarly, the outer product $\dyad{j}{\bar{j}}$ is mapped to $\dyad{0}{1}^{\otimes n}$ if $h_j < n/2$ and $\dyad{1}{0}^{\otimes n}$ if $h_j > n/2$. We again define $E$ to be the error correction matrix, define $E^{(j,k)}$ to be the entry of $E$ in the $j$th row and $k$th column, it follows from the above that
\begin{equation}
E^{(j,k)} = \begin{cases}
1, \text{ if } h_j=0 \text{ and } h_k < \frac{n}{2}\\
1, \text{ if } h_j=n \text{ and } h_k > \frac{n}{2}\\
0, \text{ otherwise}
\end{cases}.
\end{equation}

Using the same evolution described in Appendix A (thus no ancilla), we obtain that the amplitudes of the final quantum state are given by
\begin{equation}
\vec{a} = \Big( E e^{A \tau} \Big)^{t/\tau} \vec{a}_0,
\end{equation}
and,
\begin{equation}
\vec{b} = \Big( E e^{B \tau} \Big)^{t/\tau} \vec{b}_0.
\end{equation}
The solution of $\vec{a}$ is trivial where the only non-zero entries are the first and last, both of which are equal to $1/2$. The solution for $\vec{b}$ is more complicated, but the only significant terms of the matrix are the four corner entries. By discarding the other entries we obtain a reduced version of the problem
\begin{equation}
\vec{b}^\prime = \begin{pmatrix}
\eta_- & \zeta_+ \\
\zeta_- & \eta_+
\end{pmatrix}^{t/\tau} \begin{pmatrix}
1/2 \\
1/2
\end{pmatrix}
\end{equation}
where the first and second entry of $\vec{b}^\prime$ corresponds to the amplitudes of $\dyad{0}{1}^{\otimes n}$ and $\dyad{1}{0}^{\otimes n}$ respectively, and
\begin{equation}
\eta_\pm = e^{-n \gamma t}\sum_{m=0}^{\lfloor n/2 \rfloor} \binom{n}{m} x_\pm ^{n-m} y^m, \;\;\;\;\;\;\; \zeta_\pm = e^{-n \gamma t} \sum_{m=0}^{\lfloor n/2 \rfloor} \binom{n}{m} x_\pm ^{m} y^{n-m}.
\end{equation}

Notice that $\zeta_\pm \in \mathcal{O} \big( \tau^\frac{n+1}{2} \big)$, therefore
\begin{equation}
\begin{pmatrix}
\eta_- & \zeta_+ \\
\zeta_- & \eta_+
\end{pmatrix}^{t/\tau} = \begin{pmatrix}
\eta_- +\zeta_- & 0 \\
0 & \eta_+ + \zeta_+
\end{pmatrix}^{t/\tau} + \mathcal{O}\big((\gamma \tau )^\frac{n-1}{2} \big).
\end{equation}
Additionally, we have that $\big(r e^{\pm i \phi} \big)^n=\eta_\pm + \zeta_\pm$. So it follows that for large $n$ the final quantum state in this scenario is very similar to the final quantum state using the parity check code (with a noiseless ancilla and perfect error correction). Mathematically, it is equivalent up to order $(\gamma \tau )^{\frac{n-1}{2}}$, and therefore the QFI is similarly equivalent up to the same order.

\section{Fisher Information versus Quantum Fisher Information}

In the main text, we note that the QFI is inversely proportional to the minimum uncertainty of the estimate \cite{helstrom1976, braunstein1994}. The minimization is taken over all estimation strategies. The Fisher information bounds the uncertainty of the estimate for a specific information. Because of the added specificity, it can be a more meaningful quantity. For a positive operator value measure (POVM), $\{ E_j \}$, the associated Fisher information can be computed via
\begin{equation}
\label{eq:FisherInf}
    \mathcal{I} = \sum_j \frac{\Tr \big( E_j \dot{\rho} \big)^2}{\Tr \big( E_j \rho \big)}.
\end{equation}

Recall that the quantum state after perfect error correction with a noiseless ancilla is of the form
\begin{equation}
\rho = \frac{1+R}{2}\dyad*{\psi_+} + \frac{1-R}{2} \dyad*{\psi_-},
\end{equation}
with
\begin{equation}
\ket*{\psi_\pm}= \frac{1}{\sqrt{2}} \big(\ket{0}^{\otimes n+1} \pm e^{i \theta} \ket{1}^{\otimes n+1} \big).
\end{equation}
The measurement strategy involves measuring in the basis spanned the quantum states $\{ \ket{\alpha_+}, \ket{\alpha_-} \}^{\otimes (n+1)}$, where
\begin{equation}
    \ket{\alpha_\pm} = \frac{1}{\sqrt{2}} \big( \ket{0} \pm e^{i \alpha} \ket{1} \big).
\end{equation}
Due to the symmetry of $\rho$, one only needs to consider the corresponding POVMs
\begin{equation}
    E_j = \dyad{\alpha_+}^{\otimes n+1-j} \dyad{\alpha_-}^{\otimes j},
\end{equation}
for $0 \leq j \leq n+1$.

Prior to computing the Fisher information, we note that
\begin{equation}
\Tr \big( E_j \rho \big) = \frac{1+(-1)^j R \cos \big( \theta - \alpha \big)}{2^{n+1}}.
\end{equation}
Plugging the above into Eq.~(\ref{eq:FisherInf}), one obtains
\begin{equation}
    \mathcal{I} = \sum_{j=0}^{n+1} \binom{n+1}{j} \frac{\Tr \big( E_j \dot{\rho} \big)^2}{\Tr \big( E_j \rho \big)} = \frac{\Big( \dot{R} \cos \big( \theta - \alpha \big) - R \dot{\theta}\sin \big( \theta - \alpha \big) \Big)^2}{1-R^2 \cos \big( \theta - \alpha \big)}.
\end{equation}
Observe that if $\alpha$ is chosen such that $\cos \big( \theta - \alpha \big) \approx 0$, then
\begin{equation}
    \mathcal{I} = \mathcal{Q} + \mathcal{O} \big( \tau ^2 \big).
\end{equation}
Of course, this requires exact knowledge of $\omega$ to implement perfectly, and as previously mentioned, would defeat the purpose of quantum metrology. However, with adaptive estimation strategies \cite{gill2005, fujiwara2006}, this can be effectively achieved.

\section*{References}
\bibliography{QuantMetAndECC}

\begin{thebibliography}{10}
\bibitem{note1}
To further simulate a realistic setting, one could divide $\tau$ into two
  parts: one accounting for the wait time between applications of the parity
  check code, and one accounting for the time between the syndrome measurement
  and the correction. this is beyond the scope of the current work.
\end{thebibliography}

\end{document}